\documentclass[useAMS,usenatbib]{mn2e} 
\usepackage{amssymb, amsmath, graphics} 
\usepackage{graphicx} 
\usepackage{aas_macros} 
 
\usepackage[usenames]{color}

\newcommand{\msun} {{\rm M}_\odot}    
\newcommand{\cesam} {{\sc cesam}}     
\newcommand{\filou} {{\sc filou}}     
\newcommand{\amlt} {\alpha_{\rm MLT}} 
 
\title[Regularities in frequency spacings]{Regularities in frequency spacings of 
$\delta$ Scuti stars: The {\it Kepler} star KIC\,9700322\thanks{Based on observations 
obtained with the Hobby-Eberly Telescope, which is a joint project of the 
University of Texas at Austin, the Pennsylvania State University, Stanford 
University, Ludwig-Maximilians-Universit\"at M\"unchen, and Georg-August-Universit\"at G\"ottingen.}} 
 
\author[Michel Breger et al.] 
{M.~Breger$^{1,2}$, L.~Balona$^{3}$, P. Lenz$^{4,1}$, J. K. Hollek$^2$, D. W. Kurtz$^{6}$, 
\newauthor{G. Catanzaro$^{7}$, M. Marconi$^8$, A. A. Pamyatnykh$^{4,5}$, B. Smalley$^{9}$, J.C. Su\'arez$^{10}$,}
\newauthor{R. Szabo$^{11}$, K. Uytterhoeven$^{12}$, V. Ripepi$^{8}$, J. Christensen-Dalsgaard$^{13}$,}
\newauthor{H. Kjeldsen$^{13}$, M. N. Fanelli$^{14}$, K. A. Ibrahim$^{15}$, K. Uddin$^{15}$}\\
\\
$^1$Institut f\"ur Astronomie der Universit\"at Wien, T\"urkenschanzstr. 17, A--1180, Wien, Austria\\ 
$^2$Department of Astronomy, University of Texas, Austin, TX 78712, USA\\ 
$^3$South African Astronomical Observatory, P.O. Box 9, Observatory 7935, South Africa\\ 
$^4$Copernicus Astronomical Center, Bartycka 18, 00-716 Warsaw, Poland\\
$^5$Institute of Astronomy, Russian Academy of Sciences, Pyatnitskaya 48, 109017 Moscow, Russia\\
$^6$Jeremiah Horrocks Institute of Astrophysics, University of Central Lancashire, Preston PR1 2HE, UK\\
$^7$INAF - Osservatorio Astrofisico di Catania, via S. Sofia 78, 95123 Catania, Italy\\
$^8$INAF-Osservatorio Astronomico di Capodimonte, Via Moiariello 16, 80131 Napoli, Italy\\
$^9$Astrophysics Group, Keele University, Staffordshire ST5 5BG, UK\\
$^{10}$Instituto de Astrof\'{\i}sica de Andaluc\'{\i}a (CSIC), CP3004, Granada, Spain\\
$^{11}$Konkoly Observatory of the Hungarian Academy of Sciences, Konkoly Thege Mikl\'os út 15-17, H-1121 Budapest, Hungary\\
$^{12}$Laboratoire AIM, CEA/DSM-CNRS-Universit\'e Paris Diderot; CEA, IRFU, SAp, centre de Saclay, 91191, Gif-sur-Yvette, France\\
$^{13}$Department of Physics and Astronomy, Building 1520, Aarhus University, 8000 Aarhus C, Denmark\\
$^{14}$Bay Area Environmental Research Inst./NASA Ames Research Center, Moffett Field, CA 94035, USA\\
$^{15}$Orbital Sciences Corporation/NASA Ames Research Center, Moffett Field, CA 94035, USA\\} 
 
\date{Accepted 2010 month day. 
 Received 2010 month day; 
 in original form 2010 month date} 
 
\begin{document} 
 
 
\pagerange{\pageref{firstpage}--\pageref{lastpage}} \pubyear{2010} 
\maketitle 
 
\label{firstpage} 
 
\begin{abstract} 
In the faint star KIC\,9700322 observed by the {\it Kepler} satellite, 76 
frequencies with amplitudes from 14 to 29000\,ppm were detected. The two dominant 
frequencies at 9.79 and 12.57\,d$^{-1}$ (113.3 and 145.5 $\mu$Hz), interpreted to be radial modes, are 
accompanied by a large number of combination frequencies. A small additional modulation 
with a 0.16\,d$^{-1}$ frequency is also seen; this is interpreted to be the 
rotation frequency of the star. The corresponding prediction of slow rotation is 
confirmed by a spectrum from which $v \sin i = 19 \pm 1$\,km\,s$^{-1}$ is 
obtained. The analysis of the spectrum shows that the star is one of the coolest
$\delta$\,Sct variables. We also determine T$_{\rm eff}$\,=\,6700~$\pm$~100 K and
$\log g$\,=\,3.7~$\pm$~0.1, compatible with the observed frequencies of the radial modes.
Normal solar abundances are found.
An $\ell=2$ frequency quintuplet is also detected with a frequency 
separation consistent with predictions from the measured rotation rate. A remarkable 
result is the absence of additional independent frequencies down to an amplitude 
limit near 14\,ppm, suggesting that the star is stable against most forms of 
nonradial pulsation. A low frequency peak at 2.7763\,d$^{-1}$ in KIC\,9700322 is
the frequency difference between the two dominant modes and is repeated over 
and over in various frequency combinations involving the two dominant modes. The 
relative phases of the combination frequencies show a strong correlation with 
frequency, but the physical significance of this result is not clear.

 \end{abstract} 
 
 
\section{Introduction} 
 
The {\it Kepler Mission} is designed to detect Earth-like planets around solar-type 
stars \citep{Koch2010}. To achieve that goal, {\it Kepler} is continuously 
monitoring the brightness of over 150\,000 stars for at least 3.5\,yr in a 105 
square degree fixed field of view. Photometric results show that after one year of 
almost continuous observations, pulsation amplitudes of 5\,ppm are easily 
detected in the periodogram for stars brighter than $V = 10$\,mag, while at $V = 
14$\,mag the amplitude limit is about 30\,ppm. Two modes of observation are 
available: long cadence (29.4-min exposures) and short-cadence (1-min exposures). 
With short-cadence exposures \citep{gilliland2010}
it is possible to observe the whole frequency range seen in $\delta$\,Sct stars. 
 
Many hundreds of $\delta$\,Sct stars have now been detected in {\it Kepler} 
short-cadence observations. This is an extremely valuable homogeneous data set which 
allows for the exploration of effects never seen from the ground. Ground-based 
observations of $\delta$\,Sct stars have long indicated that the many observed 
frequencies, which typically span the range $5-50$\,d$^{-1}$, are mostly $p$~modes 
driven by the $\kappa$-mechanism operating in the He\,\textsc{ii} ionization zone. 
The closely-related $\gamma$\,Dor stars lie on the cool side of the $\delta$\,Sct 
instability strip and have frequencies below about 5\,d$^{-1}$. These are 
$g$~modes driven by the convection-blocking mechanism. Several stars exhibit 
frequencies in both the $\delta$\,Sct and $\gamma$\,Dor ranges and are known as 
hybrids. \cite{Dupret2005} have discussed how the $\kappa$ and convective blocking 
mechanisms can work together to drive the pulsations seen in the hybrids. 
 
The nice separation in frequencies between $\delta$\,Sct and $\gamma$\,Dor stars 
disappears as the amplitude limit is lowered. {\it Kepler} observations have shown 
that frequencies in both the $\delta$\,Sct and $\gamma$\,Dor regions are present 
in almost all of the stars in the $\delta$\,Sct instability strip 
\citep{Grigahcene2010}. In other words, practically all stars in the $\delta$\,Sct 
instability strip are hybrids when the photometric detection level is sufficiently 
low. 
 
\begin{figure*} 
\centering 
\includegraphics[bb=120 200 720 550,width=14cm]{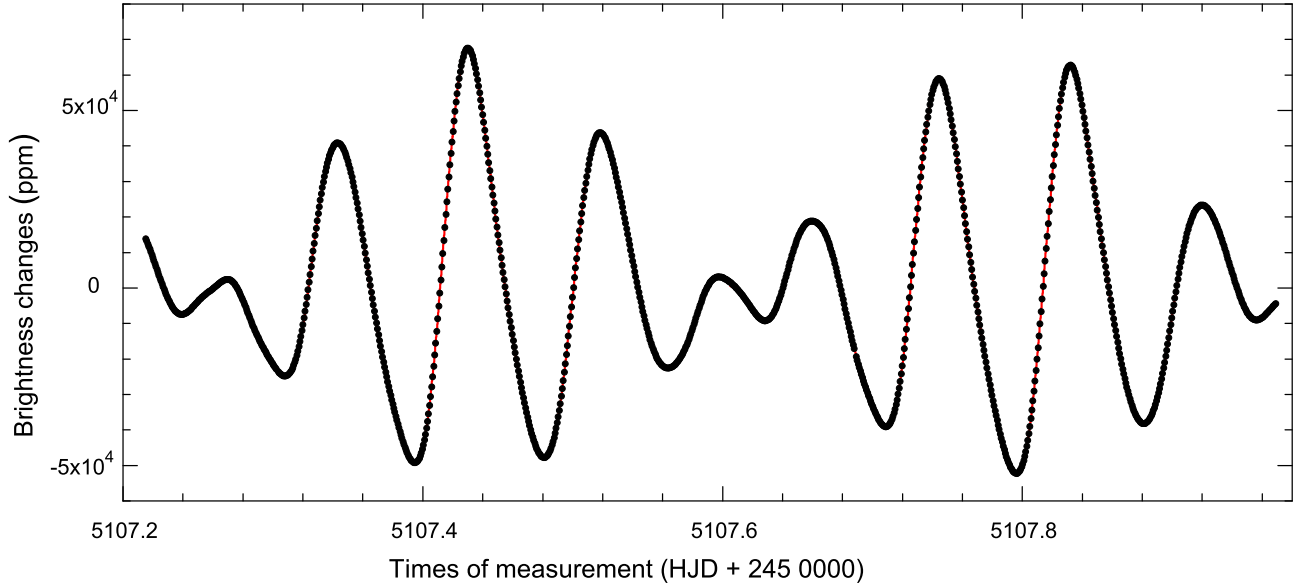} 
\caption{A sample of the {\it Kepler} light curve covering 0.8\,d. The 
multifrequency fit is shown as a solid curve. The pattern shown here roughly 
repeats every 0.72\,d.}
\label{lcurve} 
\end{figure*} 

\begin{figure*} 
\centering 
\includegraphics[width=8cm]{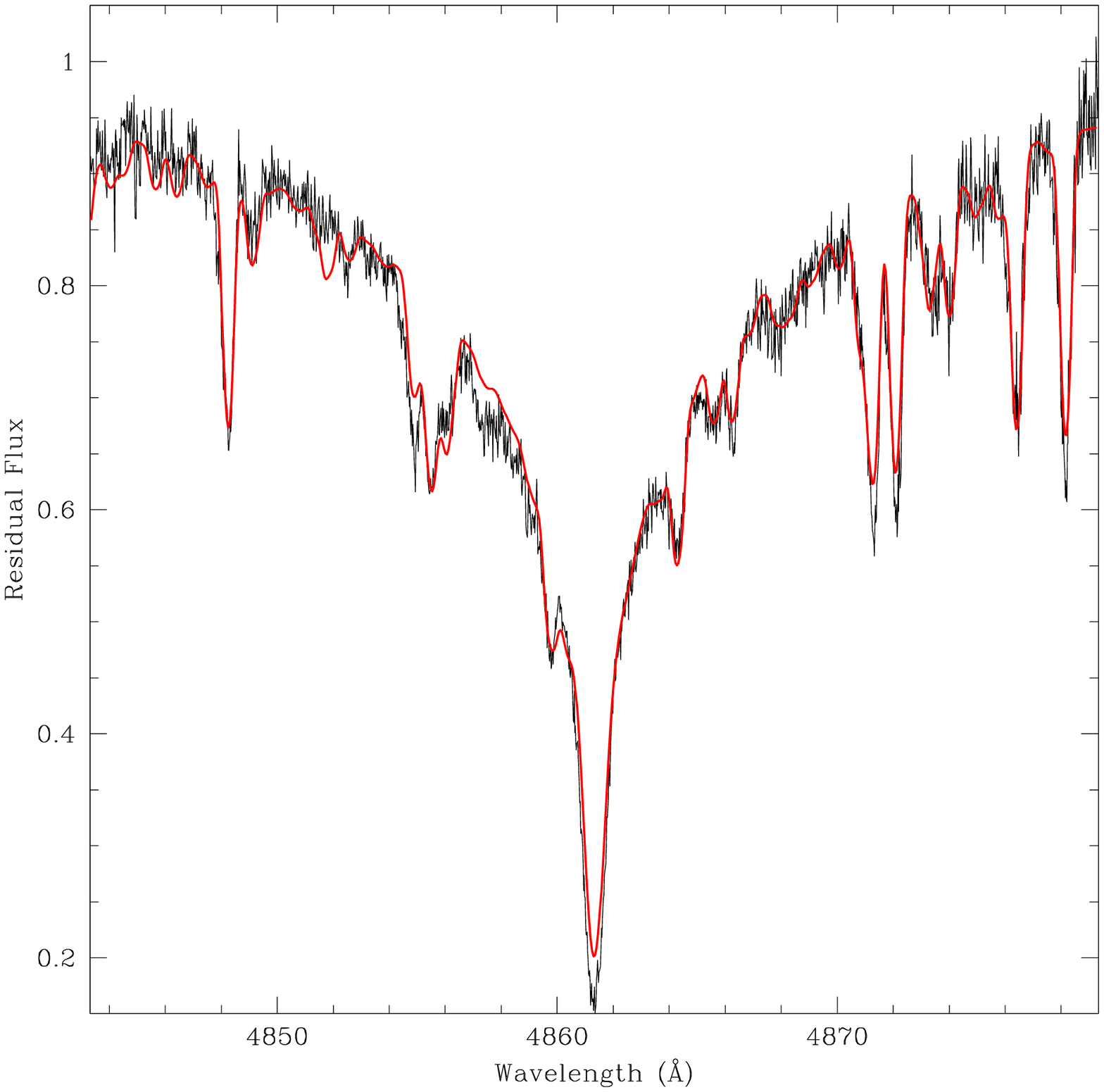} 
\includegraphics[width=8cm]{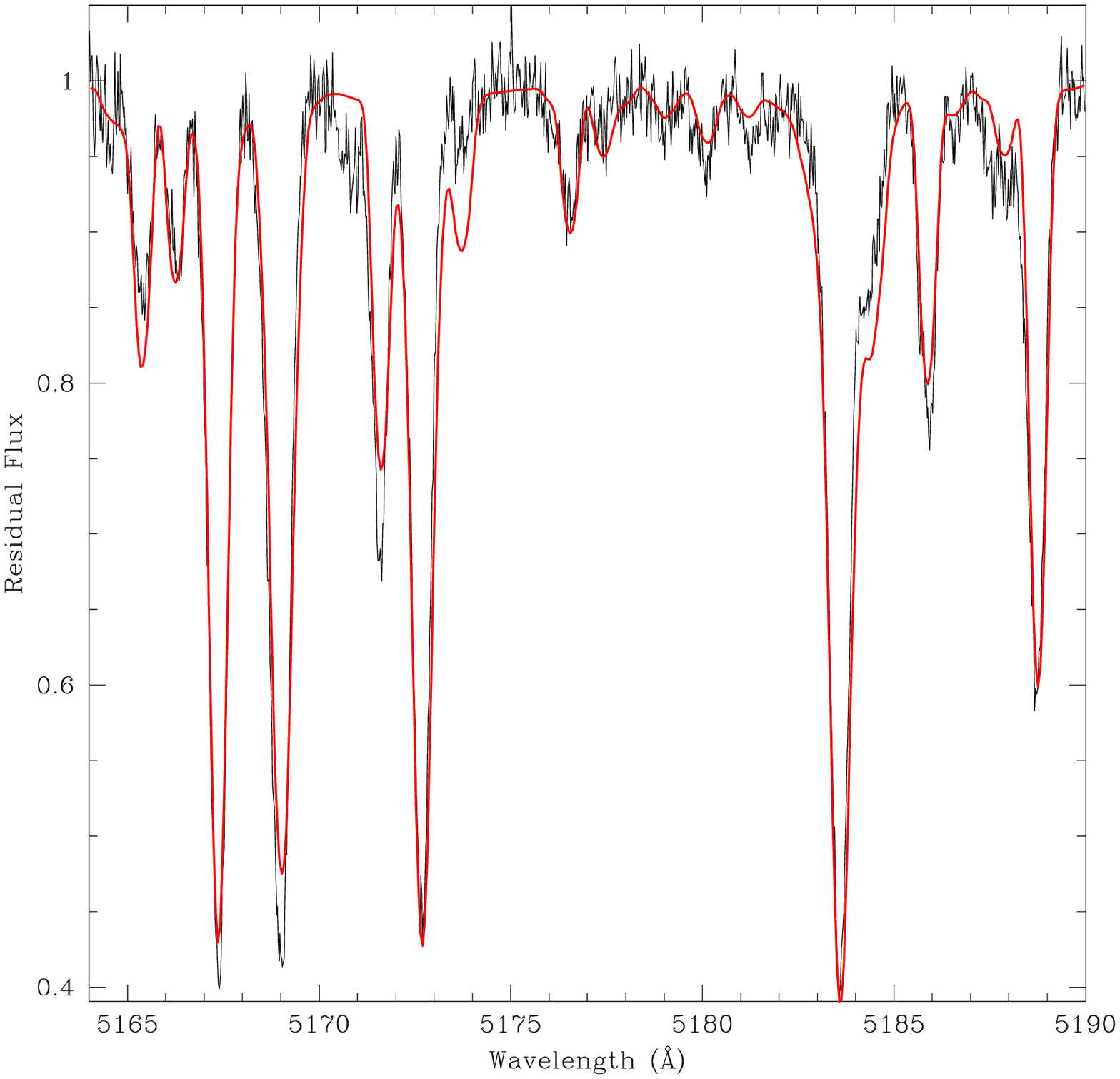} 
\caption{Two portions of observed spectrum matched to a model with 
$T_{\rm eff}$\,=\,6700, $\log g$\,=\,3.7 (red line). In the left panel we 
show the region around H$\beta$ that is sensitive to temperature, and in the
right panel the region around Mg{\sc i} triplet sensitive to gravity.} 
\label{spectrum} 
\end{figure*}

Statistical analyses of several $\delta$\,Sct stars observed from the ground have 
already shown that the photometrically observed frequencies are not distributed at 
random, but that the excited nonradial modes cluster around the frequencies of 
the radial modes over many radial orders. The observed regularities can be partly 
explained by modes trapped in the stellar envelope 
\citep{BregerLenzPamyatnykh2009}. This leads to regularities in the observed 
frequency spectra, but not to exact equidistance. 
 
In examining the {\it Kepler} data for $\delta$\,Sct stars we noticed several 
stars in which many exactly equally-spaced frequency components are present. There 
are natural explanations for nearly equally spaced frequency multiplets such as 
harmonics and non-linear combination frequencies. In some of these stars, however, 
these mechanisms do not explain the spacings. In these stars there is often more 
than one exact frequency spacing and these are interleaved in a way which
so far defies any explanation. 
 
Some examples of equally-spaced frequency components which remain unexplained are 
known from ground-based observations. The $\delta$\,Sct star 1\,Mon has a 
frequency triplet where the departure from equidistance is extremely small: only 
$0.000079 \pm 0.000001$\,d$^{-1}$ (or $0.91 \pm 0.01$\,nHz), yet the frequency 
splitting cannot be due to rotation because $\ell = 0$ for the central component 
and $\ell = 1$ for the other two modes 
\citep{BregerKolenberg2006,BalonaBartlett2001}. In the $\beta$\,Cep star 12\,Lac, 
there is a triplet with side peaks spaced by 0.1558 and 0.1553 d$^{-1}$. The 
probability that this is a chance occurrence is very small, yet photometric mode 
identification shows that two of these modes are $\ell = 2$ and the third is $\ell 
= 1$. This is therefore not a rotationally split triplet either 
\citep{Handler2006}. 
 
One solution to these puzzling equally-spaced frequencies could be non-linear mode 
interaction through frequency locking. \cite{Buchler1997} show that frequency 
locking within a rotationally split multiplet of a rapidly rotating star (150 to 200 km\,s$^{-1}$)
could yield equally-spaced frequency splitting, which is to be contrasted to the 
prediction of linear theory where strong departures from equal splitting are 
expected. 
 
In this paper we present a study of the $\delta$\,Sct star KIC\,9700322 (RA = 
19:07:51, Dec = 46:29:12 J2000, Kp = 12.685). There are two modes with 
amplitudes exceeding 20000\,ppm and several more larger than 1000\,ppm. The equal 
frequency spacing is already evident in these large amplitude modes. This star 
does not fall in the unexplained category discussed above. It is, however, a 
remarkable example of a star in which combination frequencies are dominant.

The star has a large pulsational amplitude which can easily be observed from the ground.
It was found to be variable in the "All Sky Automated Survey" \citep{Pigulski2008},
where it is given the designation
ASAS 190751+4629.2.  It is classified as a periodic variable (PER) with a
frequency of 7.79 d$^{-1}$.  This is the 2 d$^{-1}$ alias of the
main frequency (9.79 d$^{-1}$), which is determined below from the {\it Kepler} data.
The {\it Kepler} data is, of course, not affected by daily aliasing. It was also examined
during the "Northern Sky Variability Survey" \citep{Wozniak2004} with up to two measurements per night.
Due to the short periods of the star, the 109 points of NSVS 5575265 were not suitable for a comparison
with our results.

\section{New observations of KIC\,9700322} 
 
This star was observed with the {\it Kepler} satellite for 30.3\,d during
quarter 3 (BJD\,$245\,5093.21  - 245\,5123.56$) with short cadence. An overview
of the {\it Kepler} Science Processing Pipeline can be found in \citet{Jenkins2010}.
The field crowding factor given in the KAC is 0.016, which is about the average for the {\it Kepler} field.
The data were filtered by us for obvious outliers. After prewhitening the dominant modes, a number of 
additional points were rejected with a four-sigma filter as determined from the final multifrequency solution. 42990 out of 43103 data 
points could be used. We emphasize that most rejected points are extreme outliers and that the present
conclusions do not change if no editing is performed. As can be expected from near-continuous set of observations
with one measurement per minute, the spectral window is very clean with the second highest peak at
0.046 d$^{-1}$ and a height of 22\% relative to the main peak.
 
A small, typical sample of the {\it Kepler} measurements is shown in 
Fig.\,\ref{lcurve}. Inspection of the whole light curve indicates that the pattern 
shown in Fig.\,\ref{lcurve} is repeated every 0.72\,d. The repetition, however, is 
not perfect. This simple inspection already suggests, but does not prove, that 
most of the variability is caused by a few dominant modes and that additional, 
more complex effects are also present. 
 
The \citep{Wozniak2004} Input 
Catalogue also does not list any photometry for this star, but some information on the
spectral energy distribution is available. The spectral energy distribution was 
constructed using literature photometry: 2MASS \citep{Skrutskie2006}, GSC2.3 $B$ 
and $R$ \citep{Laskeretal2008}, TASS $V$ and $I$ \citep{Droegeetal2006},
and CMC14 $r'$ \citep{Evansetal2002} magnitudes. Interstellar Na D
lines present in the spectrum have equivalent widths of 60 $\pm$ 15~{m\AA} and
115 $\pm$ 20~{m\AA} for the D$_1$ and D$_2$ lines, respectively. The calibration
of \cite{MunariZwitter1997} gives $E(B-V)$= 0.03~$\pm$~0.01.

The dereddened spectral energy distribution was fitted using solar-composition
\citep{Kurucz1993} model fluxes. The model fluxes were convolved with photometric
filter response functions. A weighted Levenberg-Marquardt non-linear least-squares 
fitting procedure was used to find the solution that mimimized the difference 
between the observed and model fluxes. Since the surface gravity is poorly 
constrained by our spectral energy distribution, fits were performed for 
$\log g = 4.5$ and $\log g = 2$ to assess the uncertainty due to unconstrained 
$\log g$. A final value of $T_{\rm eff} = 7140 \pm 310$~K was found.  The 
uncertainties in $T_{\rm eff}$ includes the formal least-squares error and that
from the uncertainties in $E(B-V)$ and $\log g$. We note here that in the
next section values with considerably smaller uncertainties
are determined from high-dispersion spectroscopy.

\section{Characterization of the stellar atmosphere}

In order to classify the star with higher precision and to test the very low 
rotational velocity predicted by our interpretation of the pulsation spectrum in 
later sections, a high-dispersion spectrum is needed. KIC\,9700322 was observed 
on 2010 August 12 with the High Resolution Spectrograph \citep{Tull1998} on the 
Hobby-Eberly Telescope at McDonald Observatory. The spectrum was taken at 
$R \sim 30\,000$ using the 316g cross-disperser setting, spanning a wavelength 
region from $4120-7850$\AA. The exposure time was 1800 secs. A signal/noise
ratio of 194 was found at 593.6 nm. We reduced the data using standard techniques with 
{\tt IRAF}\footnotemark[2] routines in the echelle package. These included overscan 
removal, bias subtraction, flat-fielding, order extraction, and wavelength 
calibration. The cosmic ray effects were removed with the 
\textsc{\tt L.A. Cosmic} package \citep{vanDokkum2001}. 
 
\footnotetext[2] {{\tt IRAF} is distributed by the National Optical Astronomy 
Observatory, which is operated by the Association of Universities for Research in 
Astronomy, Inc., under cooperative agreement with the National Science 
Foundation.} 

The effective temperature, $T_{\rm eff}$, and surface gravity, $\log g$,
can be obtained  by minimizing the difference between the observed and synthetic 
spectra. We used a fit to the H$\beta$ line to obtain an estimate
of the effective temperature.  For stars with $T_{\rm eff} < 7000$~K the
Balmer lines are no longer sensitive to gravity, so we used the Mg{\sc i} 
triplet at 5167.321, 5172.684, and 5183.604 {\AA} for this purpose.
The goodness-of-fit parameter, $\chi^2$, is  defined as\\

$\displaystyle \chi^2 = \frac{1}{N} \sum \bigg(\frac{I_{\rm obs} - I_{\rm th}}{\delta I_{\rm obs}}\bigg)^2$,

\bigskip

\noindent
where $N$ is the total number of points and $I_{\rm obs}$ and $I_{\rm th}$ 
are the intensities of the observed and computed profiles, respectively.
$\delta I_{\rm obs}$ is the photon noise.  The error in a parameter was 
estimated by the variation required to change $\chi^2$ by unity.
The projected rotational velocity and the microturbulence were determined
by matching the metal lines in the range 5160 -- 5200 {\AA}.
 
From this procedure we obtained T$_{\rm eff}$\,=\,6700~$\pm$~100 K,
$\log g$\,=\,3.7~$\pm$~0.1, $v \sin i$\,=\,19~$\pm$~1 km\,s$^{-1}$,
$\xi$\,=\,2.0~$\pm$~0.5 km\,s$^{-1}$.
In Fig.~\ref{spectrum}, we show the match to observed spectrum.  The theoretical 
profiles were computed with {\tt SYNTHE} \citep{Kurucz1981} using {\tt ATLAS9}
atmospheric models \citep{Kurucz1993b}. The solar opacity distribution function
was used in these calculations.   The effective temperature calculated from the 
spectrum is somewhat lower than that obtained by matching the spectral energy 
distribution discussed in the previous section. The difference is within the statistical uncertainties.
We note that the spectrum showed no evidence for the presence of a companion.

Because of problems of line blending, we decided to use direct matching of 
rotationally-broadened synthetic spectra to the observations in order to 
determine the projected rotational velocity.  For this purpose, we divided 
the spectrum into several 100~{\AA} segments.  We derived the abundances in
each segment using $\chi^2$ minimization.  We used the line lists and atomic 
parameters in \citet{Kurucz1995} as updated by \citet{Castelli2004}. 

\begin{figure}
\includegraphics[width=8.4cm]{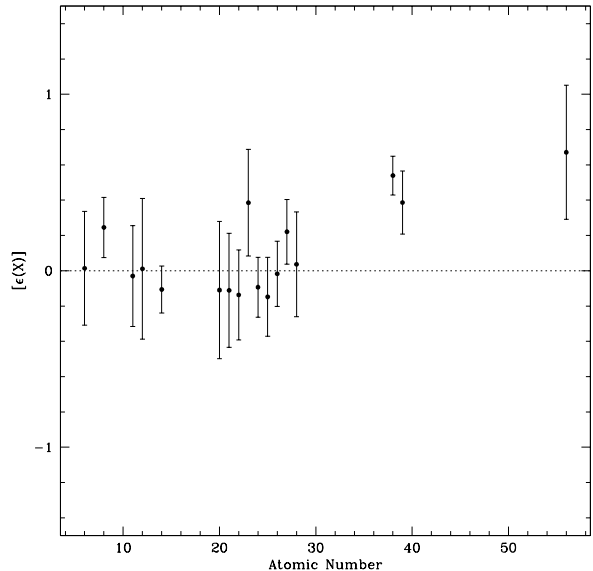}
\caption[]{Abundance pattern derived for KIC 9700322.}
\label{abundpat}
\end{figure}

Table~\ref{abund} shows the abundances expressed in the usual logarithmic 
form relative to the total number of atoms $N_{\rm tot}$. To more easily 
compare the chemical abundance pattern in KIC\,9700322,  Fig.~\ref{abundpat} 
shows the stellar abundances relative to the solar values \citep{Grevesse2010} 
as a function of atomic number.  The error in abundance for a particular element 
which is shown in Table~\ref{abund} is the standard error of the mean 
abundance computed from all the wavelength segments.  This analysis shows
that the chemical abundance in KIC\,9700322 is the normal solar abundance.

The effective temperature determined for KIC\,9700322 makes the star one of the cooler
$\delta$\,Sct stars. Other pulsators with similar temperatures are known, e.g., 6700\,K for $\rho$ Pup
\citep{netopil2008} and 6900\,K for 44 Tau \citep{LenzPamyatnykhZdravkovBreger2010}.

\begin{figure*} 
\centering 
\includegraphics[bb=20 40 800 560,width=170mm,clip]{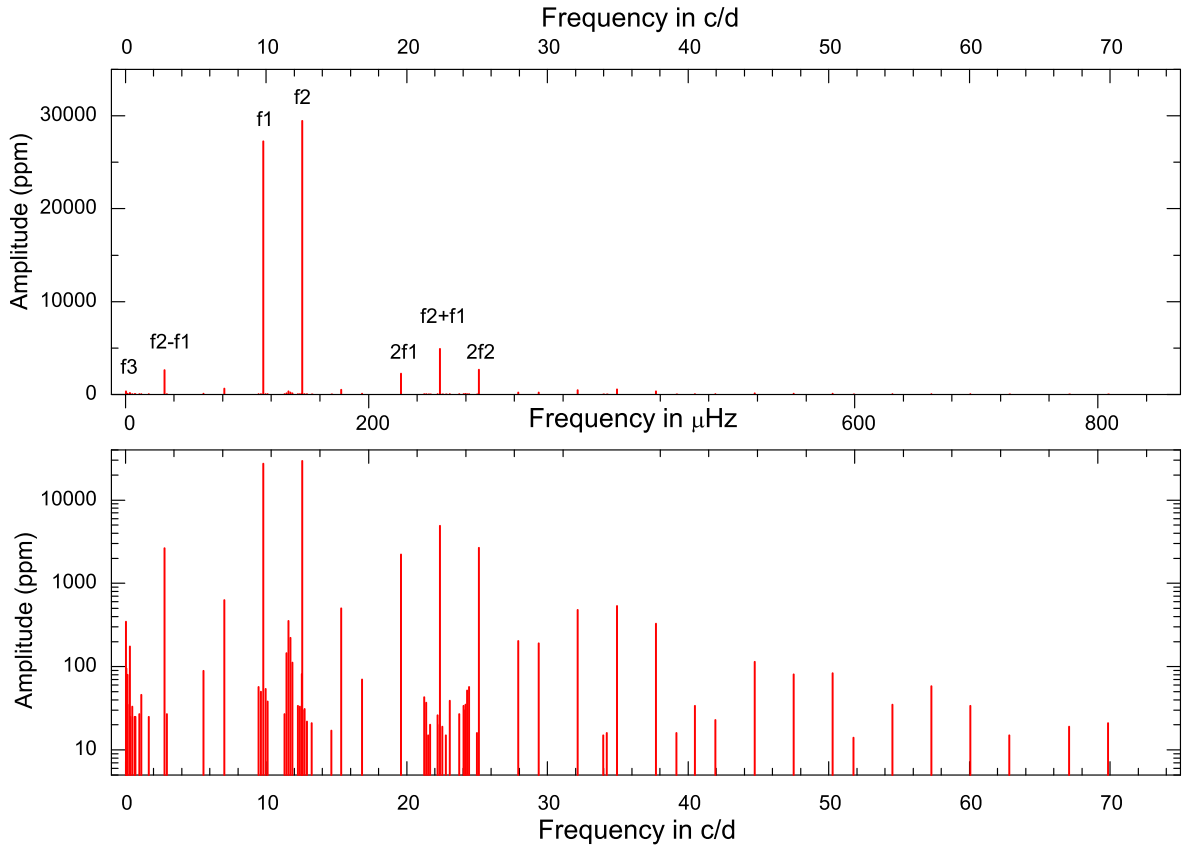} 
\caption{Amplitude spectra of KIC\,9700322. The amplitudes
(Table \ref{Table1}) are shown both linearly and logarithmically. The top panel illustrates the dominance of the two 
excited frequencies and that standard ground-based photometry would only detect 
these and a few combinations.} 
\label{pspec} 
\end{figure*}

\begin{table}
\caption[]{Abundances derived for KIC\,9700322 expressed in term of
$\log N(el)/N_{\rm tot}$}
\label{abund}
\begin{center}
\begin{tabular} {rcrcrc}
\hline
 C & $-$3.6\,$\pm$\,0.3  &   Sc & $-$9.0\,$\pm$\,0.3  &  Co & $-$6.8\,$\pm$\,0.2\\
 O & $-$3.1\,$\pm$\,0.2  &   Ti & $-$7.2\,$\pm$\,0.3  &  Ni & $-$5.8\,$\pm$\,0.3\\
Na & $-$5.8\,$\pm$\,0.3  &    V & $-$7.7\,$\pm$\,0.3  &  Sr & $-$8.6\,$\pm$\,0.1\\
Mg & $-$4.4\,$\pm$\,0.4  &   Cr & $-$6.5\,$\pm$\,0.2  &   Y & $-$9.4\,$\pm$\,0.2\\
Si & $-$4.6\,$\pm$\,0.1  &   Mn & $-$6.8\,$\pm$\,0.2  &  Ba & $-$9.2\,$\pm$\,0.4\\
Ca & $-$5.8\,$\pm$\,0.4  &   Fe & $-$4.6\,$\pm$\,0.2  &                         \\
\hline
\end{tabular}
\end{center}
\end{table}

\section{Frequency analysis} 
 
The {\it Kepler} data of KIC\,9700322 were analyzed with the statistical package {\tt 
PERIOD04} \citep{LenzBreger2005}. This package carries out multifrequency analyses 
with Fourier as well as least-squares algorithms and does not rely on the 
assumption of white noise. Previous comparisons of multifrequency analyses of 
satellite data with other techniques such as {\tt SIGSPEC} \citep{Reegen2007} have 
shown that {\tt PERIOD04} is more conservative in assigning statistical 
significances, leads to fewer \citep{Poretti2009}, and hopefully also fewer 
erroneous, pulsation frequencies, but may consequently also miss some valid 
frequencies. 
 
We did not concern ourselves with small instrumental zero-point changes in the 
data since we have no method to separate these from intrinsic pulsation. 
Consequently, our solution contains several low frequencies in the region below 
1\,d$^{-1}$ which may only be mathematical artefacts of instrumental effects. The
suspicion concerning the unreliable low frequencies is confirmed when comparing the
present {\tt PERIOD04} results with those from other period search programs
and different data editing.

Following the standard procedures for examining the peaks with {\tt PERIOD04}, we 
have determined the amplitude signal/noise values for every promising peak in the 
amplitude spectrum and adopted a limit of S/N of 3.5. The value of 3.5 (rather than 4) 
could be adopted because most low peaks do not have random frequency values due to 
their origin as combinations. This standard technique is modified for all our 
analyses of accurate satellite photometry: the noise is calculated from 
prewhitened data because of the huge range in amplitudes of three orders of 
magnitudes. 
 
After prewhitening 76 frequencies, the average residual per point was 430 ppm. 
The large number of measurements (42990)
lead to very low noise levels in the Fourier diagrams as computed by {\tt PERIOD04}:
7\,ppm ($0-10$\,d$^{-1}$), 4.7\,ppm ($10-20$\,d$^{-1}$), 
3.9\,ppm ($20-40$\,d$^{-1}$), and 3.6\,ppm ($40 - 200$\,d$^{-1}$). At low 
frequencies the assumption of white noise is not realistic. 
 
\begin{table*} 
\caption[]{Multifrequency solution of KIC\,9700322 and identifications.  
Frequencies are given in cycles d$^{-1}$ and also in $\mu$Hz.  Amplitudes 
are in parts per million (ppm).} 
\label{Table1} 
\begin{flushleft} 
\begin{tabular}{rrrclcccrc} 
\hline 
\noalign{\smallskip} 
\multicolumn{2}{c}{Frequency} & Amplitude & 	Identification & 
Comment&\multicolumn{2}{c}{Frequency} & Amplitude & 	Identification\\ 
d$^{-1}$& $\mu$Hz &ppm&&&d$^{-1}$& $\mu$Hz &ppm\\ 
\noalign{\smallskip} 
\hline 
\noalign{\smallskip} 
$\pm$ 0.0002$^1$ & $\pm$ 0.002$^1$ & $\pm$ 3$^2$\\
\noalign{\smallskip}
\multicolumn{2}{l}{Main frequencies}\\	 
\noalign{\smallskip} 
  9.7925	&	113.339	&	27266	&	$f_1$	&	Dominant mode\\			 
 12.5688	&	145.472	&	29463	&	$f_2$	&	Dominant mode\\			 
 0.1597 	&	1.848	&	80	&	$f_3$	&	 \multicolumn{2}{l}{Rotation, causes 
combinations}\\			 
 11.3163	&	130.975	&	27	&	$f_4$	&	Quintuplet	\\		 
 11.4561	&	132.593	&	145	&	$f_5$	&	Quintuplet	\\		 
 11.5940	&	134.190	&	354	&	$f_6$	&	Quintuplet	\\		 
 11.7200	&	135.648	&	221	&	$f_7$	&	Quintuplet	\\		 
 11.8593	&	137.261	&	112	&	$f_8$	&	Quintuplet	\\				
			 
\noalign{\smallskip} 
\multicolumn{3}{l}{Combination frequencies}\\								
		 
\noalign{\smallskip}	 
0.3194	&	3.697	&	174	&	$2f_3$	&	\hspace{30mm}	&	22.7723
	&	263.588	&	15	&	$f_4+f_5$	\\ 
0.4791	&	5.545	&	33	&	$3f_3$	&	\hspace{30mm}	&	23.0501
	&	266.719	&	39	&	$f_5+f_6$	\\ 
0.6388	&	7.394	&	25	&	$4f_3$	&	\hspace{30mm}	&	23.7187
	&	274.522	&	27	&	$2f_8$	\\ 
0.7095	&	8.211	&	25	&	$f_2-f_8$	&	\hspace{30mm}	&	24.0249
	&	278.066	&	34	&	$f_5+f_2$	\\ 
0.9748	&	11.282	&	27	&	$f_2-f_6$	&	\hspace{30mm}	&
	24.1628	&	279.662	&	35	&	$f_6+f_2$	\\ 
1.1127	&	12.879	&	46	&	$f_2-f_5$	&	\hspace{30mm}	&
	24.2888	&	281.120	&	52	&	$f_7+f_2$	\\ 
1.6636	&	19.254	&	25	&	$f_5-f_1$	&	\hspace{30mm}	&
	24.4282	&	282.733	&	57	&	$f_8+f_2$	\\ 
2.7763	&	32.133	&	2633	&	$f_2-f_1$	&	\hspace{30mm}	&
	24.9779	&	289.096	&	16	&	$2f_2-f_3$	\\ 
2.9360	&	33.981	&	27	&	$f_2-f_1+f_3$	&	\hspace{30mm}	&
	25.1376	&	290.945	&	2663	&	$2f_2$	\\ 
5.5526	&	64.266	&	89	&	$2f_2-2f_1$	&	\hspace{30mm}	&
	27.9139	&	323.078	&	203	&	$3f_2-f_1$	\\ 
7.0162	&	81.206	&	632	&	$2f_1-f_2$	&	\hspace{30mm}	&
	29.3776	&	340.018	&	191	&	$3f_1$	\\ 
9.4731	&	109.643	&	57	&	$f_1-2f_3$	&	\hspace{30mm}	&
	32.1538	&	372.151	&	479	&	$2f_1+f_2$	\\ 
9.6328	&	111.491	&	50	&	$f_1-f_3$	&	\hspace{30mm}	&
	33.9554	&	393.002	&	15	&	$f_6+f_1+f_2$	\\ 
9.9522	&	115.188	&	54	&	$f_1+f_3$	&	\hspace{30mm}	&
	34.2207	&	396.073	&	16	&       $f_8+f_1+f_2$	\\ 
10.1119	&	117.036	&	38	&	$f_1+2f_3$	&	\hspace{30mm}	&
	34.9301	&	404.284	&	536	&	$f_1+2f_2$	\\ 
12.2494	&	141.775	&	34	&	$f_2-2f_3$	&	\hspace{30mm}	&
	37.7064	&	436.417	&	329	&	$3f_2$	\\ 
12.4091	&	143.624	&	33	&	$f_2-f_3$	&	\hspace{30mm}	&
	39.1701	&	453.357	&	16	&	$4f_1$	\\ 
12.7285	&	147.321	&	31	&	$f_2+f_3$	&	\hspace{30mm}	&
	40.4827	&	468.550	&	34	&	$4f_2-f_1$	\\ 
12.8882	&	149.169	&	22	&	$f_2+2f_3$	&	\hspace{30mm}	&
	41.9464	&	485.490	&	23	&	$3f_1+f_2$	\\ 
15.3451	&	177.605	&	502	&	$2f_2-f_1$	&	\hspace{30mm}	&
	44.7227	&	517.623	&	114	&	$2f_1+2f_2$	\\ 
16.8088	&	194.546	&	70	&	$3f_1-f_2$	&	\hspace{30mm}	&
	47.4989	&	549.756	&	81	&	$f_1+3f_2$	\\ 
19.5850	&	226.679	&	2225	&	$2f_1$	&	\hspace{30mm}	        &
	50.2752	&	581.889	&	83	&	$4f_2$	\\ 
21.2486	&	245.933	&	43	&	$f_5+f_1$	&	\hspace{30mm}	&
	54.5152	&	630.963	&	35	&	$3f_1+2f_2$	\\ 
21.3865	&	247.529	&	37	&	$f_6+f_1$	&	\hspace{30mm}	&
	57.2915	&	663.096	&	58	&	$2f_1+3f_2$	\\ 
21.5125	&	248.987	&	15	&	$f_7+f_1$	&	\hspace{30mm}	&
	60.0678	&	695.229	&	34	&	$f_1+4f_2$	\\ 
21.6519	&	250.600	&	20	&	$f_8+f_1$	&	\hspace{30mm}	&
	62.8440	&	727.362	&	15	&	$5f_2$	\\ 
22.2016	&	256.963	&	26	&	$f_1+f_2-f_3$	&	\hspace{30mm}	&
	67.0840	&	776.435	&	19	&	$3f_1+3f_2$	\\ 
22.3613	&	258.812	&	4902	&	$f_1+f_2$	&	\hspace{30mm}	&
	69.8603	&	808.568	&	21	&	$2f_1+4f_2$	\\ 
22.5210	&	260.660	&	19	&	$f_1+f_2+f_3$	\\					
	 
\noalign{\smallskip}													
				 
\multicolumn{5}{l}{Other peaks in the amplitude spectrum}\\					
												 
\noalign{\smallskip}													
				 
0.0221	&	0.256	&	347	&	$f_{66}$	&	\hspace{30mm}	&	13.2417
	&	153.260	&	21	&	$f_{72}$	\\ 
0.0555	&	0.642	&	95	&	$f_{67}$	&	\hspace{30mm}	&	14.6254
	&	169.275	&	17	&	$f_{73}$	\\ 
0.1346	&	1.558	&	35	&	$f_{68}$	&	$f_4$-$f_5$ and $f_5$-$f_6$?
	&	22.3250	&	258.391	&	20	&	$f_{74}$	\\ 
0.3542	&	4.100	&	25	&	$f_{69}$	&	\hspace{30mm}	&	24.1464
	&	279.473	&	30	&	$f_{75}$	\\ 
12.5347	&	145.077	&	82	&	$f_{70}$	&	\hspace{30mm}	&
	51.7521	&	598.982	&	14	&	$f_{76}$	\\
12.5837	&	145.645	&	30	&	$f_{71}$\\ 	
\noalign{\smallskip} \\
\hline 
\end{tabular}
\newline
$^1$ Accuracy of frequencies determined experimentally (see Section 4.1), independent of
amplitude. The numbers
apply only to unblended frequency peaks. Because of the high quality of the $Kepler$ data, the frequency accuracy is much better than the resolution
calculated from the length of a 30.3 d run.\\
$^2$  Determined by a multiple-frequency least-squares solution.
\end{flushleft} 
\end{table*} 
 
Our analysis was performed using intensity units (ppm). The analysis was repeated
with the logarithmic units of magnitudes, which are commonly used in astronomy.
The differences in the results were,
as expected, minor and have no astrophysical implications. The only small
difference beyond the scaling factor of 1.0857 involved neighboring peaks with
large intensity differences, in which the weaker peak was in an extended 'wing' of the
dominant peak:  the effects are numerical from the multiple-least-squares solutions.

KIC\,9700322 shows only six frequencies with amplitudes larger than 1000\,ppm, of 
which only the two main frequencies are independent. Although a few ground-based 
campaigns lasting several years have succeeded in detecting statistically 
significant modes with smaller amplitudes, 1000\,ppm can be regarded as a good 
general limit. Observed with standard ground-based techniques, the star would show 
few frequencies. In all, we find 76 statistically significant frequencies.
 
\begin{figure} 
\centering 
\includegraphics[bb=30 100 750 1050,width=84mm,clip]{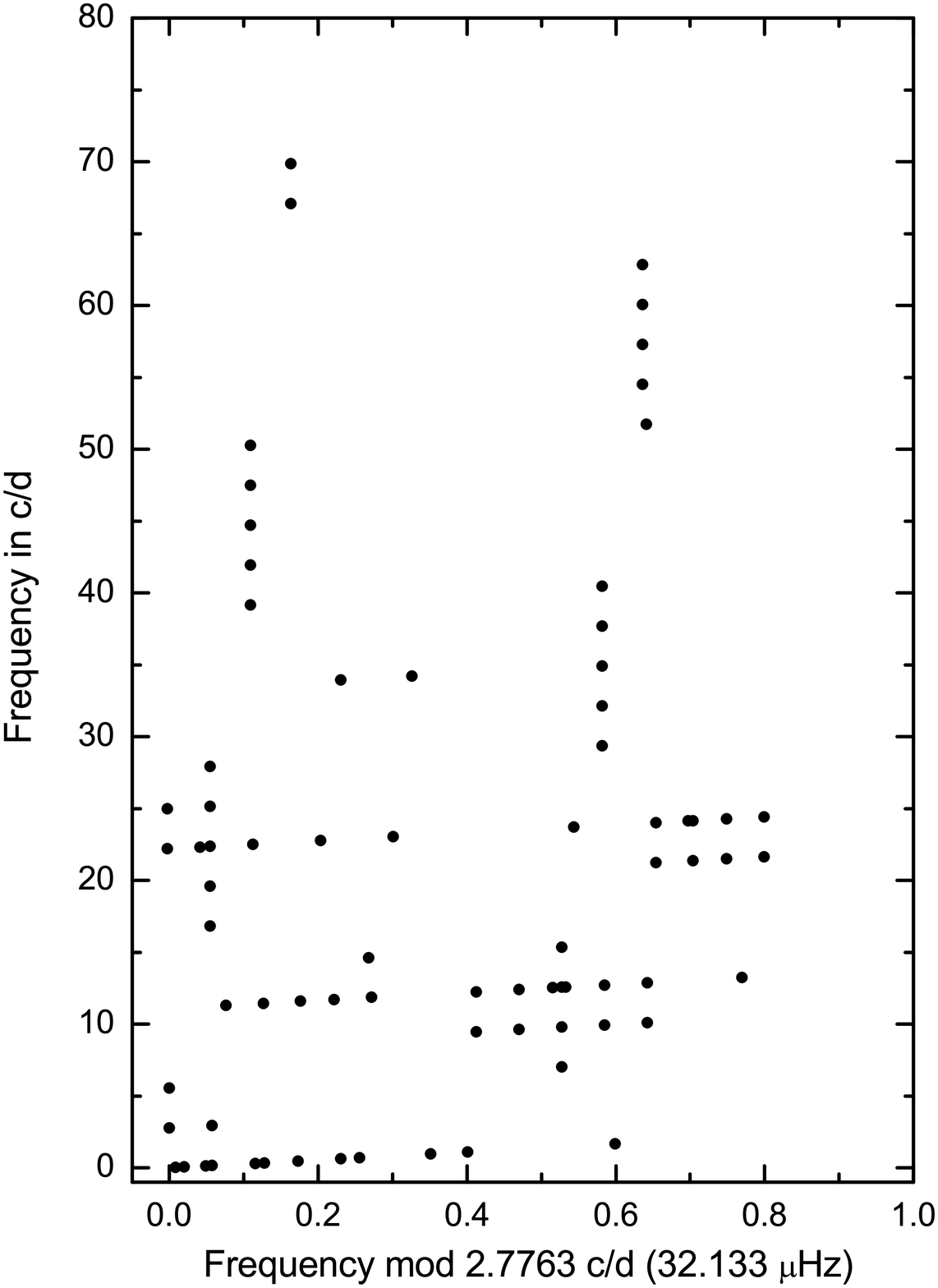} 
\caption{Echelle diagram of the detected frequencies using 2.7763\,d$^{-1}$, 
which is the difference between the two dominant pulsation modes. All patterns can 
be easily explained through combination modes, rotational splitting and modulation (see text).}
\label{echelle} 
\end{figure} 
 
\subsection{The observed frequency combinations} 
 
Most of the detected frequencies can be identified as parts of regular patterns 
(see Fig.\,\ref{pattern}). Visual inspection shows that the most obvious pattern is the exact spacing of 
$\delta f = 2.7763$\,d$^{-1}$. This is confirmed by statistical analyses of all possible frequency differences
present in the data. However, this pattern is not continued over the 
whole spectrum, but is present as different patterns, repeated and interleaved 
several times. Consequently, a simple explanation in terms of a Fourier series 
(e.g., of a nonsinusoidal light curve) is not applicable. 
 
Fig.\,\ref{echelle} shows the Echelle diagram using 2.7763$\,$d$^{-1}$, which 
demonstrates the presence of remarkable patterns. Investigation of these patterns 
reveals that they originate in very simple frequency combinations and that the 
2.7763$\,$d$^{-1}$ is only a marker of the true explanation: combinations of the 
two dominant modes $f_1$ and $f_2$, as shown in Table\,\ref{Table1}. In fact, 
2.7763\,d$^{-1}$ = ($f_2-f_1)$.

The frequencies shown in the top panel of Fig.\,\ref{pattern} can be expressed in 
as very simple way through the equation $f= m f_1\pm nf_2$, where $m$ and $n$ are 
small integers. The fact that $f_1$ and $f_2$ are the two modes with the highest 
amplitudes makes this approach also physically reasonable (see below). 
 
We also detect a frequency at 0.1597\,d$^{-1}$ (called $f_3$). This frequency is 
important, since additional patterns are also seen: a number of peaks are separated by 
exactly the value of $f_3$ (see middle panel of Fig.\,\ref{pattern}). 
 
Altogether, 57 frequencies can be identified as numerical combinations and 
multiples involving $f_1$, $f_2$ and $f_3$ by comparing the observed to the 
predicted frequencies. We can essentially rule out accidental agreements. Let us 
consider the combination frequencies at frequencies larger than 3\,d$^{-1}$, where 
the noise figures in the amplitude spectrum are reliable. For our identifications 
the average deviation between the observed and predicted frequency value is only 
0.00021\,d$^{-1}$. Such agreement is remarkable if one considers that the {\it 
Kepler} measurements used a time base of only 30\,d and that $1/T = 
0.03$\,d$^{-1}$ where $T$ is the time span between the last and first observation. 
The present result is typical for {\it Kepler} satellite data. 
 
If we use the least-squares frequency uncertainties calculated by {\tt PERIOD04}, 
on average the observed agreement is 44$\%$ {\emph{better}} than predicted. 
However, such calculations assume white noise, which is not warranted. We can 
adopt the formulae given in \cite{KallingerReegenWeiss2008} for the upper limit of 
the frequency uncertainty to include frequency-dependent noise. We calculated 
signal/noise ratios in 5\,d$^{-1}$ bins centred on each frequency with {\tt 
PERIOD04} using the prewhitened spectrum. With this more realistic approach, the 
observed deviation of 0.00021\,d$^{-1}$ is exactly a factor of two lower than the 
statistical upper limit. This supports our identifications. 
 
\subsection{The quintuplet} 
 
Five almost equidistant frequencies in the $11-12$\,d$^{-1}$ range are also 
present together with various combinations of these frequencies with $f_1$ and 
$f_2$. This is shown in the bottom panel of Fig.\,\ref{pattern}. 
 
\subsection{Explanation of the Echelle diagram} 
 
We can now explain the patterns seen in the Echelle diagram (Fig.\,\ref{echelle}) 
in a simple manner. The vertical structures are the combination frequencies 
involving $f_1$ and/or $f_2$. They are displaced from each other because different 
low integers of $m$ and $n$ (in the equation $f = m f_1\pm nf_2$) are involved. 
The horizontal structures with a slight incline correspond to the frequencies 
separated by $\sim$0.13 and 0.16\,d$^{-1}$, which are connected with the rotational
frequency of 0.1597\,d$^{-1}$ through rotational splitting and modulation.
The incline occurs because the small
frequency differences between adjacent frequency values must show up in both
the $x$ and $y$ directions of the diagram. Details on the values and 
identifications of the individual frequencies are listed in Table\,\ref{Table1}. 
 
\subsection{Additional frequencies} 
 
A few additional peaks have been identified, which are not related to $f_1 - f_8$ 
in an obvious or unique manner. The lowest frequencies were already discussed earlier 
as probable zero-point drift and their values were dependent on how the data were reduced.
The non-combination frequency at 
51.75\,d$^{-1}$ has an amplitude of only 14\,ppm. Calculation of the noise around 
the frequency gives a signal/noise ratio of 3.6, which makes it a very marginal 
detection. 
 
\begin{figure} 
\centering 
\includegraphics[bb=20 170 790 550,width=84mm,clip]{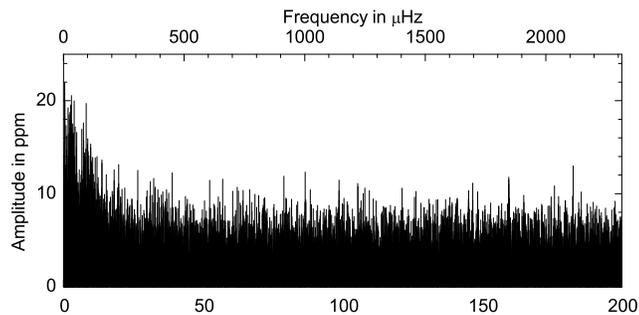} 
\caption{Amplitude spectrum of KIC\,9700322 after prewhitening 76 frequencies. 
Note the low amplitudes of any additional pulsation modes present in the star.} 
\label{prew} 
\end{figure}

\begin{figure*} 
\centering 
\includegraphics[bb=20 17 800 560,width=170mm,clip]{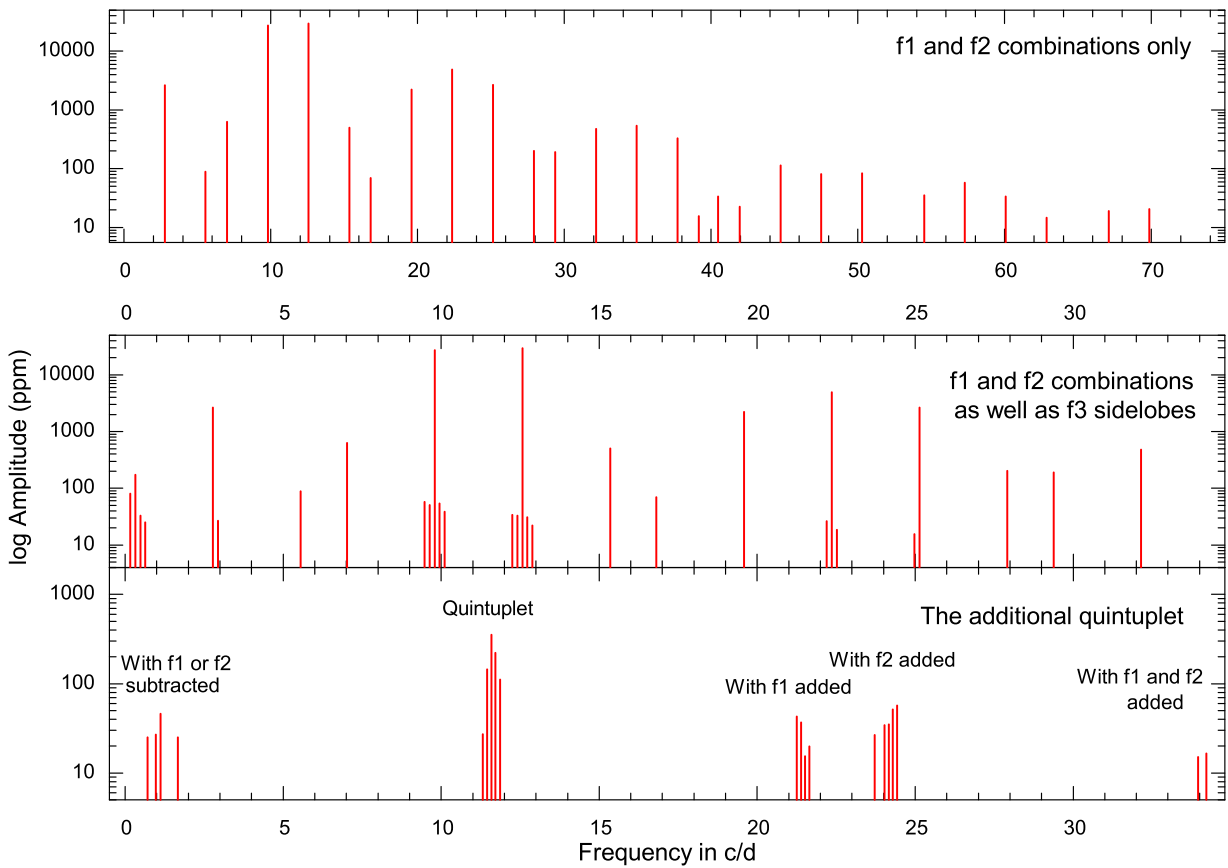} 
\caption{Top panel: Amplitude spectrum of those frequencies which are numerical 
combinations of the two dominant modes only. This demonstrates the richness of the 
combinations. Notice that the patterns are more complex than a simple Fourier sum. 
The middle panel demonstrates that some of those combination frequencies show 
further splittings with $f_3$ (rotation). The bottom panel shows an additional set of five 
frequencies together with the combinations of the set with $f_1$ and $f_2$.
} 

\label{pattern} 
\end{figure*}

Fig.\,\ref{prew} shows the amplitude spectrum of the residuals after prewhitening of the 76 
frequencies. No peak is statistically significant and the 
overall distribution of amplitudes is typical of noise. Nevertheless, we have 
examined the highest (not significant) peaks, since a few of these may be real. 
Three peaks can be identified with expected values of additional combination 
frequencies, e.g., a peak at 8.129\,d$^{-1}$ can be fit by 2$f_1$-$f_5$ at an 
amplitude signal/noise ratio of 3.0. 
 
\begin{table} 
\caption[]{The additional $l$=2 quintuplet: Observed separations} 
\label{quin} 
\begin{flushleft} 
\begin{tabular}{rr} 
\hline 
\noalign{\smallskip} 
Frequency&Separation from central frequency\\ 
d$^{-1}$&d$^{-1}$\\ 
\hline 
\noalign{\smallskip} 
$\pm$ 0.0002 & $\pm$ 0.0003\\
\noalign{\smallskip} 
11.3163	&	-0.2777	\\ 
11.4561	&	-0.1379	\\ 
11.5940	&	0	\\ 
11.7200	&	0.1260	\\ 
11.8593	&	0.2653	\\								 
\noalign{\smallskip} 
\hline 
\end{tabular} 
\end{flushleft} 
\end{table} 
 
\section{Discussion} 
 
Although this star was selected because of its very clear exactly equal frequency 
spacing, it turns out that the frequency spacing is explained as simple 
combination frequencies arising from non-linearities of the oscillation. This is 
different from another class of $\delta$\,Sct stars in the {\it Kepler} database 
which also show exact frequency spacings, but in a manner which is not at present 
understood. Examples of this strange class will be presented in a separate paper. 
 
What makes KIC\,9700322 interesting is the remarkable way in which the large 
number of frequencies are related to the two main frequencies, $f_1$ and $f_2$. 
This behaviour is very similar to the high amplitude $\delta$\,Sct star 
KIC\,9408694, also discovered in the {\it Kepler} database. The frequency patterns 
together with their amplitudes permit us to identify the different frequencies and 
to provide physical interpretations. 
 
\subsection{The dominant radial modes} 
\label{sec:radmodes}
 
The period ratio of $f_1$ and $f_2$ is 0.779. This is close to the 
expected period ratio for fundamental and first overtone radial pulsation. The 
pulsation amplitudes of $\delta$\,Sct stars increase with decreasing rotation, e.g., see Fig.\,5 of 
\cite{Breger2000}. Furthermore, high amplitudes occur mainly in slowly rotating, radial pulsators: in
fact, the high-amplitude $\delta$\,Sct (HADS) subgroup is defined on the basis of
peak-to-peak amplitudes in excess of 0.3 mag. Nevertheless, a rigid separation between radial
HADS and lower-amplitude nonradial $\delta$\,Sct stars does not exist. Dominant radial modes
with amplitudes smaller than 0.3 mag have previously been found. Examples of EE Cam \citep{bregerrucinskireegen2007}
and 44 Tau \citep{LenzPamyatnykhZdravkovBreger2010}. The situation might be summarized as follows:
Dominant radial modes occur only in slowly rotating stars.

Since KIC\,9700322 is sharp-lined ($v \sin i = 
19$\,km\,s$^{-1}$) and presumably also a slow rotator, it follows this relationship.
The presence of two dominant radial modes with amplitudes less than the
HADS limits of peak-to-peak amplitudes of 0.3 mag is not unusual.
  
The measured $v \sin i$ value supports the interpretation of the observed 
0.1597\,d$^{-1}$ peak as the rotational frequency. In fact, both dominant modes
have very weak side lobes with spacings of exactly the rotational frequency.
The side lobes are very weak: for $f_1$ and $f_2$ the amplitudes are only 0.0018 and 0.0011 of the central 
peak amplitudes. We interpret this as a very small modulation of the amplitudes with rotation.
An alternate explanation in terms of rotational splitting of nonradial modes is improbable because
rotational splitting does not lead to exact frequency separation
unless there is frequency locking due to resonance. Also, the extreme amplitude ratios
tend to favour the interpretation in terms of amplitude modulation.

Based on this mode identification assumption we investigated representative
asteroseismic models of the star. We have used two independent numerical packages: the first package consisted of the current versions of the Warsaw-New Jersey stellar evolution code and the Dziembowski pulsation code \citep{wd1977,wdgd1992}. The second package is composed by the evolutionary code \cesam\ \citep{Morel97}, and the oscillation code \filou\ \citep{Sua02thesis, Sua06rotcel}.
Both pulsation codes consider second-order effects of rotation including near degeneracy effects.

The period ratio between the first radial overtone and fundamental mode mainly 
depends on metallicity, rotation and stellar mass. Moreover, the radial period 
ratio also allows for inferences on Rosseland mean opacities as shown in 
\cite{LenzPamyatnykhZdravkovBreger2010}. 

Indeed, an attempt to reproduce the radial fundamental and first overtone mode 
at the observed frequencies with the first modelling package revealed a strong
dependence on the choice of the chemical composition and the OPAL vs. OP opacity data
\citep{IglesiasRogers1996,Seaton2005}.
The best model found in this investigation was obtained with OP 
opacities and increased helium and metal abundances. Unfortunately, this model
($T_{\rm  eff} = 7400$\,K, $\log L/{\rm L}_\odot = 1.27$, $\log g = 3.87$, 
2M$_\odot$) is much hotter than observations indicate. The disagreement in 
effective temperature indicates that this model is not correct despite the 
good fit of the radial modes.

As an additional test, by adopting the radial linear nonadiabatic models 
developed by \citet{mp98} and \citet{m04}, we are able to reproduce the values of the two 
dominant frequencies with pulsation in the fundamental and first overtone modes,
but with a lower period ratio (0.770) than observed. The best fit 
solution obtained with these models, for an effective temperature consistent 
with the spectroscopic determination and assuming solar chemical composition, 
corresponds to: $M=1.65 M_{\odot}$, $\log L/{\rm L}_\odot = 1.1$, 
$T_{\rm  eff}=6700$ K, $\log g = 3.83$. We notice that for this combination of
stellar parameters, both the fundamental and the first overtone mode  are
unstable in  these models. Moreover, looking at the Main Sequence 
and post-Main Sequence evolutionary tracks in the gravity versus effective 
temperature plane, as reported in Fig.~4 of \citet{c10}, the solution  
$T_{\rm  eff}$ = 6700 K, $\log g = 3.83$ is consistent with a $1.65 M_{\odot}$ 
stellar mass.

However, as already noted, the period ratio
in our models is lower than the observed value.
To resolve this discrepancy, the possibility of low metallicity and rotation effects was 
examined in more detail with the second modeling package. Models between 
$T_{\rm  eff} = 6200$\,K and $8600\,$K with masses between 1.2 and 1.76$\msun$,
were found to represent a good fit of $f_1$ and $f_2$ as radial fundamental and
first overtone, respectively. The best fit with the observations was found for
$M=1.2\,\msun$ models computed with $\amlt=0.5$, ${\rm d}_{ov}=0.1$, and a 
metallicity of [Fe/H]\,=\,-0.5~dex. Such a low value for the convection 
efficiency is in good agreement with the predictions by \citet{Casas06} for 
$\delta$~Sct stars, based on their non-adiabatic asteroseismic analysis. All 
these parameters roughly match the general characteristics of the $\delta$\,Sct
stars with dominant radial modes and large amplitudes, despite being in 
the limit in metallicity. 

The $P_1/P_0$ period ratios predicted by these models (which simultaneously 
fit $P_0$) are near 0.775, which is lower than the observed 
ratio, 0.779. A period ratio of 0.775 is also obtained by adopting the radial linear
nonadiabatic models by \citet{m04} at $Z=0.006$, according to which the best fit
solution with effective temperature consistent with the spectroscopic determination,
corresponds to $Z=0.006$, $Y=0.25$, $M=1.5$, $\log L/L_{\odot}=1.04$,
$T_e=6700$ K, $\log g = 3.83$. Again the fundamental and first overtone
modes are predicted to be simultaneoulsy unstable for this parameter combination.
We explored the possibility that such a discrepancy might be 
due to rotation effects, particularly second-order distortion effects, as 
discussed by \citet{Sua06pdrotI} and \citet{Sua07pdrotII}. These investigations 
analyze theoretical Petersen Diagrams including rotation effects (Rotational 
Petersen Diagrams, hereafter RPDs), and show that $P_1/P_0$ ratios increase as 
stellar surface rotation increases. The rotation rate derived from observations 
is slightly below 25\,km\,s$^{-1}$ (see section\,\ref{sec:quintuplet}). At such 
rotation rates near degeneracy effects on the period ratio are small (less than 
0.001 in $P_1/P_0$). However, when non-spherically symmetric components of the 
centrifugal force are considered, near-degeneracy effects may be larger, around 
0.0025, causing the presence of wriggles in the RPDs
(see Fig. 5 in \citet{Sua07pdrotII} and Fig. 6 in \citet{Pamyatnykh2003}).
Such effects are even more significant for rotational velocities 
higher than 40 - 50\,km\,s$^{-1}$. Consequently, near-degeneracy effects may help 
to decrease the discrepancy between the observed period ratio and the slightly 
lower values predicted by the models.

If the star had a low metal abundance (close to Pop.~II), a detailed 
analysis of RPDs might have provided an independent estimate of the true 
rotational velocity (and thereby of the angle of inclination).  However, the
spectroscopic analysis indicates that the star has a solar abundance.
KIC 9700322 therefore represents a challenge to asteroseismic modeling, since 
it appears impossible to reproduce all observables simultaneously 
with standard models.
 
\subsection{The combination frequencies} 
 
We have already shown that the 50+ detected frequency peaks can be explained by 
simple combinations of the two dominant modes and the rotational frequency. 
Several different nonlinear mechanisms may be responsible for generating 
combination frequencies between two independent frequencies, $\nu_1$ and $\nu_2$. 
For example, any non-linear transformation, such as the dependence of emergent 
flux variation on the temperature variation ($L = \sigma T^4$) will lead to cross 
terms involving frequencies $\nu_1 + \nu_2$ and $\nu_1 - \nu_2$ and other 
combinations. The inability of the stellar medium to respond linearly to the 
pulsational wave is another example of this effect. Combination frequencies may 
also arise through resonant mode coupling when $\nu_1$ and $\nu_2$ are related in 
a simple numerical way such as $2\nu_1 \approx 3\nu_2$. 

The interest in the combination frequencies derives from the fact that their 
amplitudes and phases may allow indirect mode identification. For nonradial modes, 
some combination frequencies are not allowed depending on the parity of the modes 
\citep{Buchler1997} which could lead to useful constraints on mode identification. 
Since $f_1$ and $f_2$ in KIC\,9700322 are both presumably radial, there are no 
such constraints. 
 
The identification of $f_1$ and $f_2$ with radial modes allows us to investigate the
properties of the Fourier parameters of the combination modes with the aim to disentangle
less obvious cases and/or solutions with a smaller number of combination terms.
\cite{Buchler1997} show that a resonance of the type $f_c = n_1f_1 + n_2f_2$ leads 
to a phase $\phi_r = \phi_c - (n_1\phi_1 + n_2\phi_2)$. In the same way we may 
define the amplitude ratios $A_r = A_c/(A_1A_2)$. To investigate how $\phi_r$ and 
$A_r$ behave with frequency, we first need the best estimate of the parent 
frequencies. We obtained these by non-linear minimization of a truncated Fourier 
fit involving $f_1, f_2$ and all combination frequencies up to the 4$^{\rm th}$ 
order. The best values are $f_1 = 9.792514$ and $f_2 = 12.568811$\,d$^{-1}$. The 
resulting amplitude and phases are shown in Table\,\ref{cfreq} together with the 
values of $\phi_r$ and $A_r$. The phases were calculated relative to 
BJD\,245\,5108.3849 which corresponds to the midpoint of the observations. 
 
\begin{table} 
\caption{Best fitting amplitudes, $A_c$ (ppm), and phases $\phi_c$ (radians), for 
the parent frequencies and their combination frequencies up to fourth order. 
Relative phases, $\phi_r$ and amplitudes, $A_r$ are also shown.} 
\label{cfreq} 
\begin{flushleft} 
\begin{tabular}{rrrrrr} 
\hline 
$(n_1, n_2)$ &  $f_c$     &        $A_c$  &   $\phi_c$     &   $\phi_r$  &      
$A_r$ \\    
\hline 
 (  1, 0)  &      9.792514  &   27271 &      2.710 &             &             
\\ 
 (  0,  1) &     12.568811  &   29443 &     -1.313 &             &             
\\ 
 (  2,  0) &     19.585028  &    2225 &     -0.110 &   0.751  &    0.002553 \\ 
 (  1,  1) &     22.361325  &    4898 &      1.954 &   0.557  &    0.005619 \\ 
 ( -1,  1) &      2.776297  &    2636 &      0.826 &  -1.431  &    0.003024 \\ 
 (  0,  2) &     25.137622  &    2663 &     -2.317 &   0.310  &    0.003055 \\ 
 (  3,  0) &     29.377542  &     192 &     -2.749 &   1.684  &    0.000222 \\ 
 (  2,  1) &     32.153839  &     476 &     -0.835 &   1.339  &    0.000547 \\ 
 (  2, -1) &      7.016217  &     633 &      0.127 &  -0.324  &    0.000726 \\ 
 (  1,  2) &     34.930136  &     536 &      0.202 &   0.119  &    0.000615 \\ 
 ( -1,  2) &     15.345108  &     502 &      0.525 &  -0.418  &    0.000577 \\ 
 (  0,  3) &     37.706433  &     329 &     -0.658 &  -3.000  &    0.000377 \\ 
 (  4,  0) &     39.170056  &      12 &      0.844 &   2.567  &    0.000014 \\ 
 (  3,  1) &     41.946353  &      22 &     -2.748 &   2.999  &    0.000026 \\ 
 (  3, -1) &     16.808731  &      69 &     -2.521 &   0.598  &    0.000080 \\ 
 (  2,  2) &     44.722650  &     114 &      0.869 &  -1.924  &    0.000132 \\ 
 ( -2,  2) &      5.552594  &      87 &     -2.717 &  -0.951  &    0.000101 \\ 
 (  1,  3) &     47.498947  &      84 &     -1.941 &  -0.710  &    0.000097 \\ 
 ( -1,  3) &     27.913919  &     205 &      0.199 &   0.568  &    0.000236 \\ 
 (  0,  4) &     50.275244  &      84 &     -1.061 &  -2.088  &    0.000097 \\ 
\hline 
\end{tabular} 
\end{flushleft} 
\end{table}

\begin{figure} 
\centering 
\includegraphics[scale=0.6]{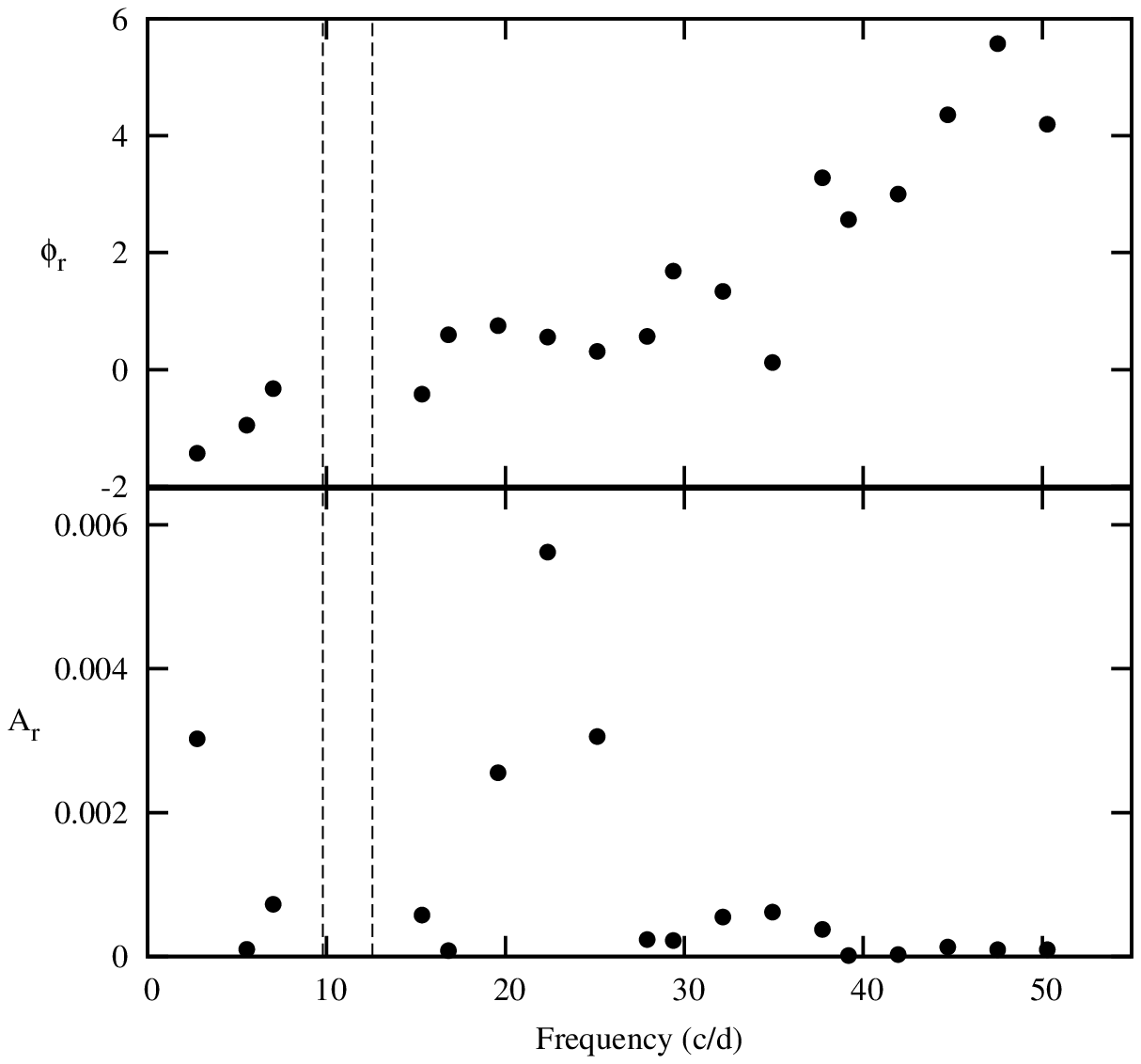} 
\caption{The relative amplitudes, $A_r$, and phases, $\phi_r$ (radians) as a 
function of frequency (d$^{-1}$).  The dotted lines show the location of the 
parent frequencies.} 
\label{comb} 
\end{figure} 
 
Fig.\,\ref{comb} shows how $A_r$ and $\phi_r$ vary with frequency. From the figure 
we note that $A_r$ is largest for $f_1 + f_2$, $2f_1$, $2f_2$ and $f_2-f_1$ and
very small for the rest. It is also interesting that $\phi_r$ is a relatively 
smooth function of frequency, being practically zero in the vicinity of the parent 
frequencies, decreasing towards smaller frequencies and increasing towards higher 
frequencies.  This result is almost independent of the choice of $f_1$ and
$f_2$.  The standard deviation of $f_1$ and $f_2$ is 0.0001~d$^{-1}$ using
the \cite{Montgomery1999} formula.  One may arbitrarily adjust $f_1$ and
$f_2$ in opposite directions by this value, and using the corresponding
calculated values of the combination frequencies, fit the data to obtain new
phases.  The resulting $\phi_r$ versus frequency remains monotonic, but the
slope does change.  The smooth monotonic nature of the $\phi_r$ versus
frequency diagram remains even for a change of ten times the standard
deviation in opposite directions for $f_1$ and $f_2$ and for arbitrary
changes in epoch of phase zero.  The result is clearly robust to
observational errors, but it is not clear what physical conclusions may be derived 
from this result. The behaviour is certainly not random and must have a physical 
basis. Note that for simple trigonometric products, $\phi_r$ will always be zero. 

Finally, we note that the amplitudes of the combination modes relative to the amplitudes of their parents
can be compared with values detected in the star 44\,Tau 
\citep{BregerLenz2008}. They agree to a factor of two or better, suggesting that 
KIC\,9700322 is not unusual in this regard, just more accurately studied because 
of the {\it Kepler} data.

\subsection{The quintuplet} 
\label{sec:quintuplet} 
 
In addition to the quintuplet structure around the two dominant modes another 
quintuplet with different properties is present in KIC 9700322 (see the 
listing of $f_4$ to $f_8$ in Table\,\ref{quin}). The average spacing between 
the frequencies in this quintuplet is slightly smaller than the rotational 
frequency (0.1338\,d$^{-1}$ vs. 0.1597\,d$^{-1}$). This makes this quintuplet 
different from the quintuplet structures found around the two dominant modes, 
which exhibit a spacing that corresponds exactly to the rotation frequency. 
Moreover, the distribution of amplitudes within the third quintuplet is 
fundamentally different to the patterns around $f_1$ and $f_2$. The given 
characteristics support an interpretation of the quintuplet as an $l$ = 2 mode.
 
The location of the quintuplet near the centre in between the radial fundamental 
and first overtone mode rules out pure acoustic character. Consequently, the 
observed quintuplet consists of mixed modes with considerable kinetic energy 
contribution from the gravity-mode cavity. For such modes theory predicts a 
smaller (and more symmetrical) rotational splittings compared to acoustic modes 
due to different values of the Ledoux constant $C_{nl}$. Using the framework of 
second order theory \citep{wdgd1992} we determined the equatorial 
rotation rate which provides the best fit of the observed quintuplet with an 
$\ell=2$ multiplet. The best results were obtained for an equatorial rotation rate 
of 23\,km\,s$^{-1}$. This is only slightly higher than the observed $v \sin i$ 
value of 19\,km\,s$^{-1}$, and therefore indicates a near-equator-on-view. The Ledoux 
constant, $C_{nl}$, of the $\ell=2$ quintuplet is 0.164. For quadrupole modes 
$C_{nl}$ ranges between $\approx$0.2 for pure gravity modes to smaller values 
for acoustic modes. With (1 - $C_{nl}$) = 0.836 this leads to a rotational frequency, 
$\nu_{rot}=\frac{\Omega}{2\Pi}$, of around 0.16 d$^{-1}$. Consequently, this theoretical result 
confirms the interpretation of $f_3$ as a rotational feature and of the quintuplet 
as $l$ = 2 modes. Further support is provided by the fact that we see various 
combinations of the quintuplet with $f_1$ and $f_2$. 

Moreover, the location of the quintuplet allows us to determine the extent of 
overshooting from the convective core. In the given model we obtained 
$\alpha_{ov}=0.13$ but the uncertainties elaborated in 
Section~\ref{sec:radmodes} currently prevent an accurate determination.
 
\subsection{Further discussion} 
 
A remarkable aspect of the star is the fact that so few pulsation modes are 
excited with amplitudes of 10\,ppm or larger. 
 
In the interior of an evolved $\delta$\,Sct star, even high-frequency $p$~modes 
behave like high-order $g$~modes. The large number of spatial oscillations of 
these modes in the deep interior leads to severe radiative damping. As a result, 
nonradial modes are increasingly damped for more massive $\delta$\,Sct stars, 
which explains why high-amplitude$\delta$\,Sct stars pulsate in mostly radial 
modes and why in even more massive classical Cepheids nonradial modes are no 
longer visible. 
 
In general, we do not expect the frequencies in the $\delta$\,Sct stars observed by 
{\it Kepler} to be regularly spaced because, unlike ground-based photometry, the 
observed pulsation modes are not limited to small spherical harmonic degree, 
$l$. For the very low amplitudes detected by {\it Kepler} we may expect to see 
a large number of small-amplitude modes with high $l$. The observed amplitudes 
decreases very slowly with $l$ and, all things being equal, a large number of 
modes with high $l$ might be expected to be seen in $\delta$\,Sct and other 
stars \citep{Balona1999}. The $\delta$\,Sct stars HD\,50844 \citep{Poretti2009} 
and HD\,174936 \citep{Hernandez2009} observed by {\it CoRoT} show many hundreds of 
closely-spaced frequencies and may be examples of high-degree modes. The relatively small 
number of independent frequencies detected in KIC\,9700322 stands in strong 
contrast to the two stars observed by {\it CoRoT}. 
 
It should be noted that, unlike many $\delta$\,Sct stars observed by {\it Kepler}, 
KIC\,9700322 does not have any frequencies in the range normally seen in 
$\gamma$\,Dor stars. The only strong frequencies in this range are a few 
combination frequencies. Although we have identified significant frequencies below 
0.5\,d$^{-1}$, it is not possible at this stage to verify whether these are due to 
the star or instrumental artefacts. At present, we do not understand why low 
frequencies are present in so many $\delta$\,Sct stars. 
 
Regularities in the frequency spacing due to combination modes have already been 
observed from the ground even in low amplitude $\delta$\,Sct stars. An example is 
the star 44\,Tau \citep{BregerLenz2008}. Fig.\,2 of \cite{BregerLenzPamyatnykh2009} 
demonstrates that all the observed regularities outside the $5-13$\,d$^{-1}$ 
range are caused by combination modes. For combination modes the frequency 
spacing must be absolutely regular within the limits of measurability. This 
is found for KIC\,9700322.

\section*{Acknowledgements}

MB is grateful to E. L. Robinson and M. Montgomery for helpful discussions. This investigation has been supported
by the Austrian Fonds zur F\"{o}rderung der wissenschaftlichen Forschung through project P 21830-N16. 
LAB which to acknowledge financial support from the South African
Astronomical Observatory.  AAP and PL acknowledge partial financial support from 
the Polish MNiSW grant No. N N203 379 636. This work has been supported
by the `Lend\"ulet' program of the Hungarian Academy of Sciences and Hungarian OTKA grant K83790.

The authors wish to thank the {\it Kepler} team for their generosity in
allowing the data to be released to the {\it Kepler} Asteroseismic Science Consortium
(KASC) ahead of public release and for their outstanding efforts which have
made these results possible.  Funding for the {\it Kepler} mission is provided
by NASA's Science Mission Directorate.

\bibliographystyle{mn2e} 
\bibliography{KIC9700322_ref2} 
 
\label{lastpage} 
 
\end{document}